\documentclass[11pt]{article}
\usepackage{graphicx}
\usepackage{amsmath,amssymb}
\def\1{{\bf 1}}

\def\be{\begin{equation}}
\def\ee{\end{equation}}

\begin{document}
\noindent {\Large \bf Molecular Dynamics Studies of the Bufallo Prion Protein Structured Region}\\

\bigskip

\noindent {\Large Jiapu Zhang$^\text{ab*}$}\\
{\small {\it
\noindent $^\text{a}$Graduate School of Sciences, Information Technology and Engineering \& Centre of Informatics and Applied Optimisation, Faculty of Science, The Federation University Australia, Mount Helen Campus, Mount Helen, Ballarat, Victoria 3353, Australia;

\noindent $^\text{b}$Molecular Model Discovery Laboratory, Department of Chemistry \& Biotechnology,  Faculty of Science, Engineering \& Technology, Swinburne University of Technology, Hawthorn Campus, Hawthorn, Victoria 3122, Australia; 

%Correspondence address: 
\noindent $^\text{*}$Tel: +61-3-9214 5596, +61-3-5327 6335; 
jiapuzhang@swin.edu.au, j.zhang@federation.edu.au\\
}}

\noindent {\bf Abstract:} 
 {\it Molecular dynamics (MD) studies of buffalo prion protein (BufPrP$^\text{C}$) (J Biomol Struct Dyn. 2016 Apr;34(4):762-77) showed that the structure of this protein is very stable at room temperature (whether under neutral pH or low pH environments).  In order to understand the reason why buffalo is resistant to prion diseases and why BufPrP$^\text{C}$ is so stable at room temperature, this paper will prolong our MD running time at room temperature and extend our research to higher temperatures to study this BufPrP$^\text{C}$ structure furthermore. We found an important reason why BufPrP$^\text{C}$ is so stable at room temperature and this might be a nice clue of drug discovery or drug design for the treatment of prion diseases.\\ 
}


\section{Introduction}
Unlike bacteria and viruses, which are based on DNA and RNA, prions are unique as disease-causing agents since they are misfolded proteins. Prions propagate by deforming harmless, correctly folded proteins into copies of themselves. The misfolding is irreversible. Prions attack the nervous system of the organism, causing an incurable, fatal deterioration of the brain and nervous system until death occurs. Some examples of these diseases are mad cow disease in cattle, chronic wasting disease in deer and elk, and Creutzfeldt-Jakob disease in humans.\\

Not every species is affected by prion diseases. Water buffalo is a species being resistant to prion diseases. The research question arises: from the molecular structure point of view, what is the reason that allows it to retain its molecular structure folding? This is the research question addressed in this paper. Many experimental studies have shown that BufPrP is very stable so that it resists to the infection of diseased prions \cite{iannuzzi_etal1998, oztabak_etal2009, imran_etsl2012, zhao_etal2012, uchida_etal2014, qing_etal2014} - the brief review of these experimental studies can be seen in \cite{zhangwc2016}. In addition, recently Zhao et al. (2015) reported that the prion protein gene polymorphisms associated with bovine spongiform encephalopathy susceptibility differ significantly between cattle and buffalo \cite{zhao_etal2015} and reported three significant findings in buffalo: 1) extraordinarily low deletion allele frequencies of the 23- and 12-bp indel polymorphisms; 2) significantly low allelic frequencies of six octarepeats in coding sequence and 3) the presence of S4R, A16V, P54S, G108S, V123M, S154N and F257L substitutions in buffalo coding sequences \cite{zhao_etal2015}.\\

As we all know, prion diseases are caused by the conversion from normal cellular prion protein (PrP$^C$) to diseased infectious prions (PrP$^{Sc}$); in structure the conversion is mainly from $\alpha$-helices to $\beta$-sheets (generally PrP$^C$ has 42\% $\alpha$-helix and 3\% $\beta$-sheet, but PrP$^{Sc}$ has 30\% $\alpha$-helix and 43\% $\beta$-sheet \cite{griffith1967, jones_etal2005, daude2004, ogayars1998, panbn1993, reilly2000}). The structural region of a PrP$^C$ usually consists of $\beta$-strand 1 ($\beta$1), $\alpha$-helix 1 (H1), $\beta$-strand 2 ($\beta$2), $\alpha$-helix 2 (H2), $\alpha$-helix 3 (H3), and the loops linked them each other]. The conformational changes may be amenable to study by molecular dynamics (MD) techniques. NMR experiences showed that a prion resistant species does not have higher conformational stability at higher temperature than nonresistant species \cite{yu_etal2016}; this means at high temperature the $\alpha$-helices of PrP$^C$ will turn to $\beta$-sheets of PrP$^{Sc}$ so that we can find out some secrets of the protein structural conformational changes of PrP. Hence, in this paper we will use MD to study the molecular structure of buffalo prion protein BufPrP$^C$(124--227) \cite{zhangwc2016}. In \cite{zhangwc2016}, the structure of BufPrP$^C$(124--227) is showed very stable at room temperature, whether under neutral pH or low pH environments. In order to understand the reason why buffalo is resistant to prion diseases and why BufPrP$^\text{C}$ is so stable at room temperature, this paper will prolong our MD running time at room temperature and extend our research to higher temperatures to study this BufPrP$^\text{C}$ structure furthermore - the Methods and Materials will be briefly given in the next section. In Section 3, we will analyze our MD computational results and discuss a reason why BufPrP$^\text{C}$ is so stable at room temperature. Section 4 presents a concluding remark of this paper and propose a nice clue from BufPrP studies for drug discovery/design of the treatment of prion diseases.

\section{Materials and Methods}
The MD structure used for the paper is the one of \cite{zhangwc2016}, i.e. the structured region BufPrP structure. In \cite{zhangwc2016}, 25 ns' MD simulations were done for room temperature 300 K. This paper prolongs the MD running time to 30 ns. Moreover, this paper extends 30 ns' MD simulations to higher temperatures 350 K and 450 K respectively. The MD methods for 350 K \cite{zhang2012} and 450 K \cite{zhang2010, zhang2011b, zhangl2011, zhangwz2015} are completely same as the ones of \cite{zhang2012, zhang2010, zhang2011b, zhangl2011, zhangwz2015, zhangBook2015}. We emphasize that all of our methods are completely reproducible \cite{chencl2016}.\\ 

BufPrP has stable molecular structures at 300 K during the whole 25 ns' MD simulations \cite{zhangwc2016}, where the root mean square deviation (RMSD) and Radius Of Gyration values are not changing very much during the whole 25 ns. As we all know, RMSD and Radius Of Gyration are two indicators for structural changes in a protein. The Radius Of Gyration is the mass weighted scalar length of each atom from the center-of-mass (COM). The RMSD is used to measure the scalar distance between atoms of the same type for two structures. In this paper, the initial structures compared with all the MD structures are the minimized/optimized structures. From the RMSD and Radius Of Gyration observations for 300 K, 350 K and 450 K, we then carry out deeper researches on the secondary structures developments during the whole 30 ns' MD simulations of 300 K, 350 K, 450 K. The BufPrP molecular structure is maintained by a network of atoms by their peptide bonds, covalent bonds (e.g. the disulfide bond S-S between Cys179 and Cys214), and noncovalent bonds such as hydrogen bonds, salt links, van der Waals contacts, and hydrophobic interactions. We will find out which bonds are contributing to maintain the stability of BufPrP.

\section{Results and Discussion}
\begin{figure}[h!] 
\centerline{
\includegraphics[width=6.6in]{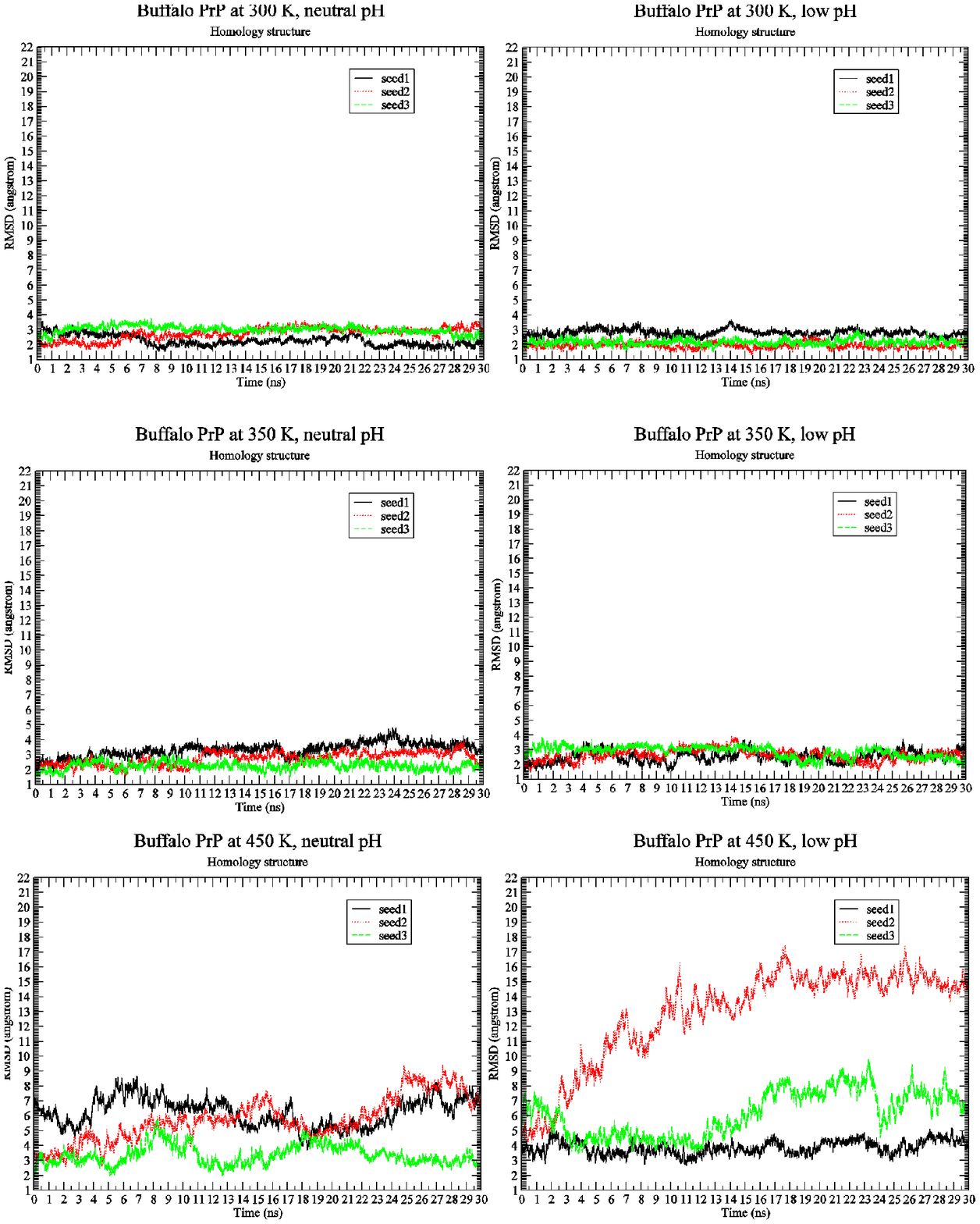}
}
\caption{\textsf{RMSD of BufPrP at 300 K, 350 K and 450 K, neutral and low pH values (up to down: 300 K, 350 K, 450 K; left: neutral pH, right: low pH) during 30 ns' MD.}}  \label{buffalo_RMSD_300K_350K_450K}
\end{figure}

\begin{figure}[h!] 
\centerline{
\includegraphics[width=6.6in]{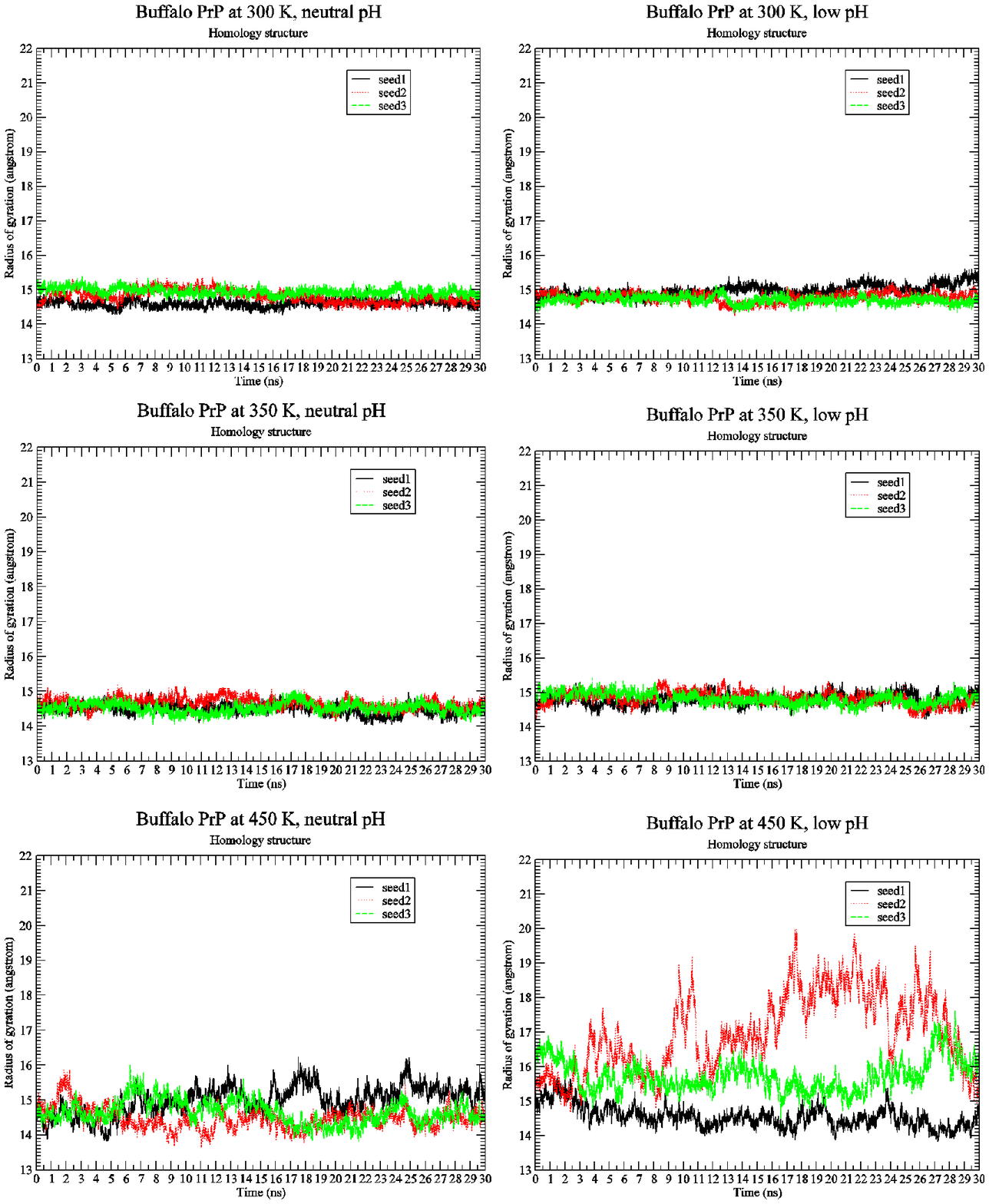}
}
\caption{\textsf{Radius Of Gyration of BufPrP at 300 K, 350 K and 450 K, neutral and low pH values (up to down: 300 K, 350 K, 450 K; left: neutral pH, right: low pH) during 30 ns' MD.}}  \label{buffalo_RadGyr_300K_350K_450K}
\end{figure}

We first present the RMSD (Figure \ref{buffalo_RMSD_300K_350K_450K}) and Radius Of Gyration (Figure \ref{buffalo_RadGyr_300K_350K_450K}) results of BufPrP at 300 K, 350 K and 450 K respectively during the whole 30 ns.

\centerline{$<$Figure 1$>$}

\centerline{$<$Figure 2$>$}

\begin{figure}[h!] 
\centerline{
\includegraphics[width=6.6in]{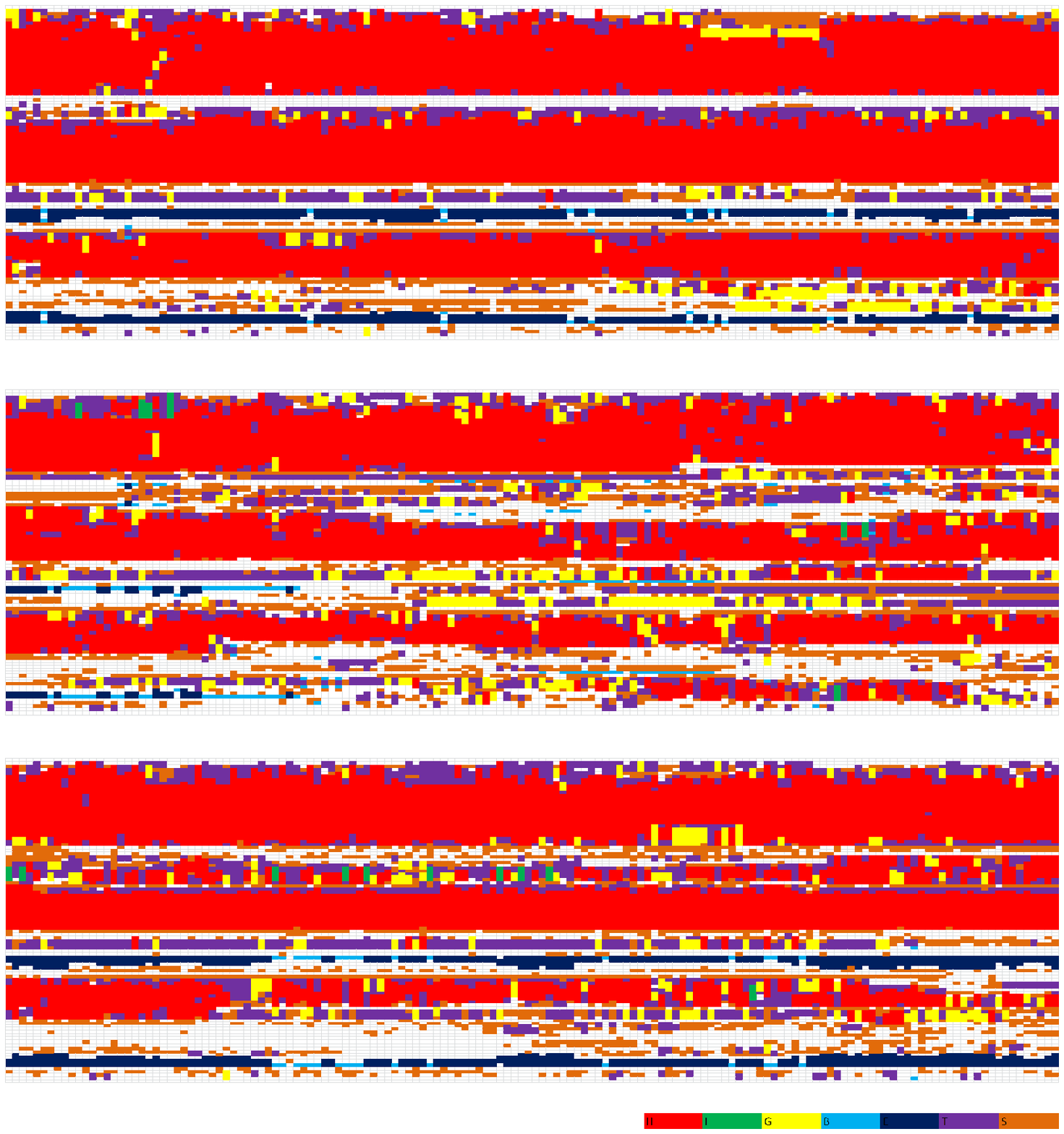}
}
\caption{\textsf{Secondary Structure graphs of BufPrP at 450 K, low pH value (up to down: seed1--seed3; X-axis: time (0--30 ns), Y-axis: residue number (124–-227); H is the $\alpha$-helix, I is the 5 helix or called $\pi$-helix, G is the 3-helix or called 3$_{10}$ helix, B is the residue in isolated $\beta$-bridge, E is the extended strand (participates in $\beta$-ladder), T is the hydrogen bonded turn, and S is the bend) during 30 ns' MD.}}  \label{buffalo_2ndaryStruct_450K_pH3}
\end{figure}

\noindent In Figures \ref{buffalo_RMSD_300K_350K_450K}$\sim$\ref{buffalo_RadGyr_300K_350K_450K}, we see that (i) at 300 K and 350 K, the values of RMSD and Radius Of Gyration are not changing very much (RMSDs varying within 2 angstroms and Radius Of Gyrations varying within 1 angstrom - these are within normal variations of MD structures because typically we would want our RMSd to be less than 1.5$\sim$2 angstroms), whether under neutral or low pH environments; (ii) the RMSD performance of (i) happens at 450 K under neutral pH environment but RMSDs varying largely nearly within 7 angstroms (however from Figure \ref{buffalo_2ndaryStruct_450K_pH3} we think the $\alpha$-helices are still unfolded under neutral pH environment at 450 K because variations of Radius Of Gyrations are normally within 2 angstroms), and the Radius Of Gyration performance of 450 K under neutral pH environment is slight worse than at 300 K and 350 K but still normally varying within 2 angstroms; and (iii) at 450 K under low pH environment, for seed2, it is clear that the $\alpha$-helices are unfolded into $\beta$-structures. Thus, next we just see the Secondary Structure graphs of BufPrP at 450 K, low pH value (Figure \ref{buffalo_2ndaryStruct_450K_pH3}).

\centerline{$<$Figure 3$>$}

\noindent Seeing Figure \ref{buffalo_2ndaryStruct_450K_pH3}, we know that not only for seed2 but also for seed3, $\alpha$-helices H1 and H2 unfold into other forms of secondary structures. In summary, in the below we may only focus on low pH environment at 450 K to find out reasons of the unfolding of $\alpha$-helices H1 and H2.\\

Because the change of pH environments from neutral pH to low pH will make the loss of salt bridges, we will mainly analyze the noncovalent bonds of salt bridges  (SBs) as follows (Tables \ref{bufPrP_SBs_300K_350K_450K}$\sim$\ref{bufPrP_SBs_300K_350K_450K_continuation}). 

%\centerline{$<$Table 1$>$}

%\centerline{$<$Table 2$>$}

\begin{table}[h!]
\caption{\textsf{Percentages (\%) of some salt bridges (between two residues) under neutral pH environment for BufPrP at 300 K, 350 K and 450 K during 30 ns' MD:}}
\centering
{\scriptsize
\begin{tabular}{l                          ||c     |c      |c        ||c      |c     |c      ||c     |c      |c} \hline 
                Buffalo PrP/               &       &300 K  &         &       &350 K  &       &       &450 K  &\\ \hline 
                Salt Bridges (SBs)         &seed1  &seed2  &seed3    &seed1  &seed2  &seed3  &seed1  &seed2  &seed3\\\hline \hline
                ASP147@CG--ARG148@CA.CZ   &100   &100    &100      &100    &100    &100    &99.95  &99.97  &99.98\\ \hline  
                HIS155@CG--ARG156@CA.CZ  &99.74 &99.80  &99.63    &99.80  &99.87  &99.75  &99.72  &99.40  &99.50\\ \hline
                HIS155@NE2--ARG156@CA.CZ &5.18  &6.53   &7.49     &31.90  &46.75  &13.26  &24.63  &18.23  &22.53\\ \hline
                GLU211@CD--ARG208@CA.CZ   &99.47 &99.70  &93.46    &91.26  &98.07  &99.67  &97.08  &97.38  &95.88\\ \hline 
                GLU207@CD--LYS204@CA.NZ   &98.48 &99.88  &99.86    &99.19  &98.05  &98.63  &87.38  &97.52  &97.67\\ \hline
                GLU221@CD--ARG220@CA.CZ   &96.78 &63.88  &52.51    &96.77  &87.26  &94.58  &69.02  &69.40  &77.25\\ \hline
                GLU186@CD--LYS185@CA.NZ   &93.63 &92.88  &96.30    &68.63  &84.71  &86.76  &19.93  &77.72  &74.98\\ \hline
                ASP178@CG--ARG164@CA.CZ   &87.89 &23.89  &1.51     &38.56  &33.93  &47.25  &4.30   &30.45  &39.87\\ \hline
                GLU196@CD--LYS194@CA.NZ   &70.57 &60.08  &19.20    &28.13  &20.43  &5.61   &4.65   &26.40  &25.12\\ \hline
                GLU207@CD--ARG208@CA.CZ   &57.47 &35.53  &73.37    &51.93  &66.71  &62.95  &81.35  &78.92  &77.83\\ \hline
                ASP147@CG--HIS140@ND1.HD1 &38.91 &20.51  &15.12    &46.31  &52.45  &64.40  &0.10   &49.72  &25.20\\ \hline
                GLU152@CD--ARG148@CA.CZ   &34.72 &23.28  &30.46    &50.64  &38.77  &40.89  &46.23  &19.42  &28.72\\ \hline 
                GLU152@CD--ARG151@CA.CZ   &33.62 &36.68  &31.59    &39.79  &34.66  &41.93  &40.73  &50.43  &37.97\\ \hline
                ASP144@CG--ARG148@CA.CZ   &27.93 &85.55  &75.43    &32.47  &52.25  &49.01  &2.48   &34.45  &52.88\\ \hline
                ASP147@CG--ARG151@CA.CZ   &19.40 &48.61  &27.63    &27.83  &20.75  &25.65  &2.98   &51.53  &51.65\\ \hline
                HIS187@NE2--ARG156@CA.CZ &14.09 &59.43  &68.19    &64.04  &18.17  &53.34  &       &21.77  &14.53\\ \hline
                HIS187@CG--ARG156@CA.CZ  &0.04  &0.13   &0.33     &5.29   &0.35   &11.79  &       &1.03   &4.23\\ \hline
                GLU221@CD--ARG164@CA.CZ   &      &38.33  &6.32     &0.02   &0.25   &0.33   &0.48   &0.33   &0.07\\ \hline
                ASP178@CG--HIS177@ND1.HD1 &13.85 &23.91  &0.40     &15.38  &14.53  &12.28  &29.15  &20.82  &21.22\\ \hline
                GLU211@CD--HIS177@ND1.HD1 &8.29  &4.67   &88.56    &23.20  &24.23  &7.75   &24.53  &8.77   &7.85\\ \hline
                GLU196@CD--ARG156@CA.CZ   &5.51  &10.04  &16.45    &65.75  &0.86   &42.63  &0.05   &28.88  &36.37\\ \hline
                GLU186@CD--HIS187@ND1.HD1 &5.13  &0.81   &0.29     &7.95   &26.29  &1.13   &80.43  &5.88   &9.35\\ \hline
                HIS187@CG--LYS185@CA.NZ  &2.03  &0.13   &0.62     &0.16   &1.18   &0.32   &0.57   &1.20   &3.63\\ \hline
                ASP202@CG--ARG156@CA.CZ   &1.69  &2.19   &21.40    &5.68   &3.33   &1.07   &0.50   &0.67   &2.10\\ \hline
                GLU207@CD--HIS177@ND1.HD1 &0.77  &1.70   &6.87     &6.35   &5.36   &1.39   &6.30   &1.32   &1.58\\ \hline
                ASP144@CG--HIS140@ND1.HD1 &0.22  &1.53   &         &0.64   &0.83   &10.14  &17.32  &10.10  &0.32\\ \hline
                HIS155@NE2--ARG136@CA.CZ &0.18  &3.15   &0.15     &0.53   &0.19   &0.61   &3.63   &3.53   &0.12\\ \hline
                HIS155@CG--ARG136@CA.CZ  &      &0.29   &         &0.05   &0.01   &0.07   &0.50   &0.88   &\\ \hline
                HIS140@NE2--ARG208@CA.CZ &0.17  &       &         &       &       &0.03   &       &4.02   &19.77\\ \hline
                HIS155@NE2--ARG151@CA.CZ &0.14  &0.83   &0.97     &       &0.10   &0.01   &38.38* &20.33  &32.00\\ \hline
                HIS155@CG--ARG151@CA.CZ  &      &0.04   &0.01     &       &0.01   &       &49.22* &18.75  &41.98\\ \hline
                GLU196@CD--HIS155@ND1.HD1 &0.06  &       &0.51     &11.34  &0.17   &6.62   &0.05   &2.35   &4.05\\ \hline
                GLU196@CD--HIS187@ND1.HD1 &      &0.03   &0.75     &0.08   &0.01   &0.15   &0.28   &1.07   &1.40\\ \hline
                HIS187@NE2--HIS155@ND1.HD1 &0.01  &       &0.05     &0.25   &0.07   &0.83   &0.02   &0.05   &\\ \hline
                GLU152@CD--HIS155@ND1.HD1 &0.01  &4.61   &         &0.42   &0.45   &0.01   &30.00* &35.52  &30.53\\ \hline
                GLU146@CD--LYS204@CA.NZ   &0.01  &       &2.18     &0.57   &0.02   &0.01   &       &       &\\ \hline
                GLU200@CD--LYS204@CA.NZ   &      &0.07   &         &       &0.03   &0.09   &2.12   &0.32   &0.23\\ \hline
                HIS155@NE2--LYS194@CA.NZ &      &       &0.88     &3.34   &2.05   &       &0.03   &1.63   &0.23\\ \hline
                GLU211@CD--HIS140@ND1.HD1 &      &       &         &0.25   &       &0.01   &       &1.93   &0.48\\ \hline
                HIS140@NE2--ARG136@CA.CZ &      &       &         &0.17   &       &       &5.18   &4.73   &5.00\\ \hline
                HIS155@CG--LYS194@CA.NZ  &      &       &         &0.16   &0.17   &       &       &0.33   &\\ \hline
                HIS140@NE2--ARG151@CA.CZ &      &       &         &0.15   &0.01   &0.03   &5.60   &1.52   &8.05\\ \hline
                HIS187@NE2--LYS194@CA.NZ &      &       &         &0.04   &       &       &1.60   &0.03   &2.50\\ \hline
                GLU146@CD--HIS140@ND1.HD1 &0.06  &       &         &       &       &       &1.93   &2.92   &22.98\\ \hline
                ASP202@CG--LYS204@CA.NZ   &      &0.01   &         &       &       &       &1.68   &0.02   &\\ \hline
\end{tabular}
} 
\label{bufPrP_SBs_300K_350K_450K}
\end{table}
\begin{table}[h!]
\caption{\textsf{Percentages (\%) of some salt bridges (between two residues) under neutral pH environment for BufPrP at 300 K, 350 K and 450 K during 30 ns' MD (continuation):}}
\centering
{\scriptsize
\begin{tabular}{l                          ||c     |c      |c        ||c      |c     |c      ||c     |c      |c} \hline 
                Buffalo PrP/               &       &300 K  &         &       &350 K  &       &       &450 K  &\\ \hline 
                Salt Bridges (SBs)         &seed1  &seed2  &seed3    &seed1  &seed2  &seed3  &seed1  &seed2  &seed3\\\hline \hline
                GLU146@CD--ARG151@CA.CZ   &      &       &         &       &       &       &25.58  &13.33  &1.33\\ \hline
                HIS140@CG--ARG151@CA.CZ  &      &       &         &       &       &       &6.47   &2.70   &4.92\\ \hline
                HIS140@CG--ARG148@CA.CZ  &      &       &         &       &       &       &0.17   &0.02   &0.08\\ \hline
                HIS140@NE2--ARG148@CA.CZ &      &       &         &       &       &       &1.18   &0.03   &0.22\\ \hline
                HIS187@CG--LYS194@CA.NZ  &      &       &         &       &       &       &1.52   &       &0.32\\ \hline
                GLU146@CD--ARG148@CA.CZ   &      &       &         &       &       &       &16.88  &3.52   &\\ \hline
                HIS140@CG--HIS155@ND1.HD1 &      &       &         &       &       &      &8.57   &0.02   &1.85\\ \hline
                HIS140@NE2--HIS155@ND1.HD1 &      &       &         &       &       &     &5.82   &0.02   &1.93\\ \hline
                HIS140@NE2--ARG156@CA.CZ &      &       &         &       &       &       &2.08   &       &\\ \hline
                HIS140@CG--ARG156@CA.CZ  &      &       &         &       &       &       &1.25   &       &\\ \hline
                HIS155@CG--HIS140@ND1.HD1 &      &       &         &       &       &      &8.03   &0.02   &1.93\\ \hline
                HIS155@NE2--HIS140@ND1.HD1 &      &       &         &       &       &     &7.78   &0.03   &2.47\\ \hline
                ASP202@CG--LYS194@CA.NZ   &      &       &         &       &       &       &6.57   &       &\\ \hline 
                GLU152@CD--ARG156@CA.CZ   &      &       &         &       &       &       &6.27   &5.30   &5.30\\ \hline 
                ASP147@CG--LYS194@CA.NZ   &      &       &         &       &       &       &5.15   &       &\\ \hline 
                ASP147@CG--LYS204@CA.NZ   &      &       &         &       &       &       &1.80   &       &\\ \hline
                ASP147@CG--ARG208@CA.CZ   &      &       &         &       &       &       &       &0.03   &\\ \hline
                GLU207@CD--LYS185@CA.NZ   &      &       &         &       &       &       &3.48   &       &0.02\\ \hline
                GLU152@CD--HIS140@ND1.HD1 &      &       &         &       &       &       &3.28   &       &\\ \hline
                GLU152@CD--ARG136@CA.CZ   &      &       &         &       &       &       &       &0.18   &\\ \hline
                GLU152@CD--LYS194@CA.NZ   &      &       &         &       &       &       &1.70   &       &0.27\\ \hline 
                GLU221@CD--ARG136@CA.NZ   &      &       &         &       &       &       &1.80   &       &\\ \hline 
                GLU221@CD--ARG136@CA.CZ   &      &       &         &       &       &       &       &0.85   &\\ \hline
                GLU221@CD--HIS140@ND1.HD1 &      &       &         &       &       &       &       &11.73  &\\ \hline
                GLU200@CD--LYS185@CA.NZ   &      &       &         &       &       &       &1.55   &       &\\ \hline
                GLU200@CD--HIS187@ND1.HD1 &      &       &         &       &       &       &1.15   &       &\\ \hline
                HIS140@NE2--ARG220@CA.CZ &      &       &         &       &       &       &0.93   &       &\\ \hline
                HIS140@CG--ARG220@CA.CZ  &      &       &         &       &       &       &0.63   &       &\\ \hline
                GLU186@CD--ARG156@CA.CZ   &      &       &         &       &       &       &0.60   &       &\\ \hline
                ASP202@CG--ARG148@CA.CZ   &      &       &         &       &       &       &0.42   &       &\\ \hline
                ASP144@CG--ARG208@CA.CZ   &      &       &         &       &       &       &0.27   &       &\\ \hline
                ASP144@CG--ARG151@CA.CZ   &      &       &         &       &       &       &0.23   &1.83   &\\ \hline
                ASP144@CG--HIS155@ND1.HD1 &      &       &         &       &       &       &0.18   &0.25   &\\ \hline
                GLU200@CD--LYS194@CA.NZ   &      &       &         &       &       &       &0.17   &       &\\ \hline
                ASP167@CG--ARG164@CA.CZ   &      &       &         &       &       &       &0.15   &       &\\ \hline
                GLU196@CD--ARG148@CA.CZ   &      &       &         &       &       &       &0.13   &       &\\ \hline
                GLU196@CD--HIS140@ND1.HD1 &      &       &         &       &       &       &0.03   &       &\\ \hline
                GLU146@CD--HIS155@ND1.HD1 &      &       &         &       &       &       &0.10   &0.08   &\\ \hline
                ASP202@CG--HIS140@ND1.HD1 &      &       &         &       &       &       &0.07   &       &\\ \hline
                HIS177@NE2--ARG208@CA.CZ &      &       &         &       &       &       &0.02   &       &\\ \hline
                HIS177@NE2--LYS185@CA.NZ &      &       &         &       &       &       &0.02   &       &\\ \hline
                HIS140@CG--ARG208@CA.CZ  &      &       &         &       &       &       &       &2.85   &6.77\\ \hline
                ASP202@CG--HIS187@ND1.HD1 &      &       &         &       &       &       &       &1.07   &3.02\\ \hline
                ASP147@CG--HIS155@ND1.HD1 &      &       &         &       &       &       &       &0.02   &0.02\\ \hline
                ASP147@CG--ARG136@CA.CZ   &      &       &         &       &       &       &       &0.22   &\\ \hline
                HIS187@NE2--LYS185@CA.NZ &      &       &         &       &       &       &1.05   &       &\\ \hline
                ASP144@CG--ARG136@CA.CZ   &      &       &         &       &       &       &       &0.03   &\\ \hline
                HIS140@NE2--LYS204@CA.NZ &      &       &         &       &       &       &       &       &0.03\\ \hline
                HIS155@NE2--ARG148@CA.CZ &      &       &         &       &       &       &0.02   &0.05   &\\ \hline
                GLU196@CD--LYS185@CA.NZ   &      &       &         &       &       &       &0.03   &       &\\ \hline
                GLU146@CD--ARG136@CA.CZ   &      &       &         &       &       &       &0.05   &6.10   &\\ \hline
                GLU146@CD--ARG208@CA.CZ   &      &       &         &       &       &       &       &4.05   &0.03\\ \hline
                HIS155@CG--HIS187@ND1.HD1 &      &       &         &       &       &      &       &       &0.02\\ \hline
                HIS140@CG--ARG136@CA.CZ  &      &       &         &       &       &       &2.45   &3.90   &1.45\\ \hline
\end{tabular}
} 
\label{bufPrP_SBs_300K_350K_450K_continuation}
\end{table}

\noindent We can see in Tables \ref{bufPrP_SBs_300K_350K_450K}$\sim$\ref{bufPrP_SBs_300K_350K_450K_continuation} the following SBs with high occupied rates at 300 K, 350 K and 450 K during the whole 30 ns of MD simulations:
\begin{enumerate}
\item[] $\bullet$ SBs in H1:
        \begin{enumerate} 
        \item[] - ASP147--ARG148, 
        \item[] - HIS155--ARG156, 
        \item[] - GLU152--ARG148, 
        \item[] - GLU152--ARG151, 
        \item[] - ASP147--ARG151,
        \item[] - ASP147--HIS140, 
        \end{enumerate}
\item[] $\bullet$ SBs in H2:
        \begin{enumerate}
        \item[] - GLU186--LYS185, 
        \item[] - HIS187--LYS185,
        \item[] - ASP178--HIS177,
        \end{enumerate}
\item[] $\bullet$ SBs in H3:
        \begin{enumerate} 
        \item[] - GLU211--ARG208, 
        \item[] - GLU207--LYS204, 
        \item[] - GLU221--ARG220, 
        \item[] - GLU207--ARG208, 
        \end{enumerate}
\item[] $\bullet$ special SBs:
        \begin{enumerate} 
        \item[] - ASP178--ARG164 - linking the $\beta$2-$\alpha$2 loop, 
        \item[] - ARG164--GLU221 - linking the $\beta$2-$\alpha$2 loop and H3,
        \item[] - GLU196--LYS194 - in the H2--H3 loop,
        \item[] - GLU196--ARG156 - linking H1 and the loop of H2--H3,
        \item[] - HIS187--ARG156 - linking H2 and the 3$_{10}$-helix after H1,
        \end{enumerate}
\end{enumerate}
\noindent and at 450 K, the SBs in H1 such as HIS155--ARG151, HIS155--GLU152 are having high occupied rates and seeing Table \ref{bufPrP_SBs_300K_350K_450K_continuation} we may know that there are many low occupied rate SBs not owned at 300 K and 350 K. The removal of all these SBs under low environment will lead to the changes of H1 and H2 region of BufPrP from $\alpha$-helices structures (BufPrP$^C$) into $\beta$-sheet structures (BufPrP$^{Sc}$). 

\section{A concluding remark}
This brief article talks about the MD results of BufPrP at different temperatures, and presents a clue to seek the reasons of the conversion from normal cellular prion protein (PrP$^C$) to diseased infectious prions (PrP$^{Sc}$). This should be very useful for the goals of medicinal chemistry in prion diseases research fields.

\section*{Acknowledgments}
This research was supported by a Melbourne Bioinformatics grant numbered FED0001 on its Peak Computing Facility at the University of Melbourne, an initiative of the Victorian Government (Australia).

\end{document}